\documentstyle[prl,aps,epsf,floats]{revtex}
\newcommand{\eq}{\begin{equation}}
\newcommand{\ee}{\end{equation}}
\newcommand{\eqa}{\begin{eqnarray}}
\newcommand{\eea}{\end{eqnarray}}

\def\sxxo{{\sigma_{xx}^0}}

\def\sxyo{{\sigma_{xy}^0}}

\def\uv{{\epsilon_{\mu\nu}}}

\def\dmu{{\partial_\mu}}
\def\dnu{{\partial_\nu}}

\def\Tr{{\rm Tr}}

\newcommand{\pprl}{Phys. Rev. Lett. \ } 
 
\begin{document}
\twocolumn[
\hsize\textwidth\columnwidth\hsize\csname@twocolumnfalse\endcsname
\draft 
]
\noindent{\bf 
Wang and Plerou Reply:} 
\ In the preceding article \cite{comment}, Tanaka and Machida (TM) comment
on the bare values of the coupling constants in an effective field theory 
derived for the transport properties in the multilayer quantum Hall 
systems \cite{wang}. 
In Ref.~\cite{wang}, the field theory was derived by
bosonization of a quantum quasi-1d fermionic theory using the
nonabelian chiral anomaly.
The latter arises following a finite chiral transformation of
Dirac fermions, and is given by the corresponding 
Fujikawa Jacobian \cite{fujikawa}. 
It was first pointed out by Roskies and Schaposnik \cite{roskies}, in the
context of the Schwinger model, that a chiral rotation with a finite angle
should be carried out by a sequence of infinitesimal rotations to 
produce the correct coupling constants in the anomalous terms.
As correctly noted by TM, this was overlooked in Ref.~\cite{wang}.
It is, however, obvious that the precise values of the bare coupling constants
are {\it totally irrelevant} for the analysis of the theory in 
Ref.~\cite{wang}, as the latter is based on the fixed point values of the 
field theory in two and three dimensions.

TM then argue in their comment \cite{comment} that the
Fujikawa Jacobian becomes zero and thus no anomaly arises
when a one parameter family of infinitesimal chiral transformations
is used to carry out the finite chiral rotation in
the nonabelian case considered in Ref.~\cite{wang}.
This claim is {\it incorrect}. Below, we will show explicitly that
the correct chiral anomaly associated with the non-invariance
of the fermion functional integral measure is obtained by
the transformations used in Ref.~\cite{wang}. We further prove
that the chiral transformation suggested by TM is
equivalent to the combined unitary and chiral rotations used in
Ref.~\cite{wang}.

To make connection to the comment, we will consider the case
of vanishing interlayer tunneling. The fermionic quantum
theory becomes that of a U(2n) Hubbard chain at half-filling.
For small hopping alternations $\delta t\ll t$, Eq.~(5) in Ref.~\cite{wang}
can be written in terms of the Dirac spinors $\Psi^{\rm T}=(\psi_R,\psi_L)$,
and ${\bar\Psi}=({\bar\psi}_R, {\bar\psi}_L)\gamma_0$,
\eq
S=\Tr {\bar \Psi}({\bf I}\otimes{\not\!\partial}+m Q
e^{-2i Q \Delta\theta\gamma^5})\Psi.
\label{sna} 
\ee
Here, $Q=u\Lambda u^\dagger$ with $u\in U(2n)$ and 
$\Lambda=\left(\begin{array}{cc}
\bf{I_{n}}& 0 \\  0 & -\bf{I_{n}} \end{array} \right),$
${\not\!\partial}=\gamma_\mu\partial_\mu$ with $\gamma_0$ and
$\gamma_1$ given by the Pauli matrices $\tau_x$ and $\tau_y$,
$\gamma^5=i\gamma_0\gamma_1$, $m=-\Delta_0$, and
$\Delta\theta=\delta t/\Delta_0$.

As in Ref.~\cite{wang}, we make a unitary transformation
${\Psi}\to u\Psi$, ${\bar\Psi}\to{\bar\Psi}u^\dagger$ in Eq.~(\ref{sna}),
leading to,
\eq
S=\Tr{\bar \Psi}({\bf I}\otimes{\not\!\partial}+i{\not\!\!A}+m
\Lambda e^{-2i\Lambda\Delta\theta\gamma^5})\Psi,
\label{sab}
\ee
where the gauge field ${\not\!\!A}=-iu^\dagger{\not\!\partial}u$.
We now carry out the sequence of infinitesmal 
chiral transformations parametrized by the parameter $t\in [0,1]$,
\eq
\Psi\to U_5(t) \Psi,\quad {\bar\Psi}\to{\bar\Psi}U_5(t),
\label{u5}
\ee
with $U_5(t)=\exp(it\Lambda\phi\gamma^5)$ and $\phi=(\Delta\theta+\pi/4)$.
The transformed action is $S=\Tr{\bar \Psi}{\not\!\!D}_t\Psi+\ln{\cal J}_F$ 
where ${\cal J}_F$ is the corresponding Jacobian and
\eq
{\not\!\!D}_t={\bf I}\otimes{\not\!\partial}+iU_5(t){\not\!\!A}U_5(t)
-im\gamma^5e^{2i\Lambda\phi(t-1)\gamma^5}.
\label{d}
\ee
The finite chiral rotation in Eq.~(6) of Ref.~\cite{wang} is built up
by iterating Eq.~(\ref{u5}). The chiral anomaly arises from the
accumulated Jacobian of the transformations and is given by \cite{sch},
\eq
\ln {\cal J}_F
=-{1\over2\pi}\int_0^1 dt(i\phi\Lambda\gamma^5){\not\!\!D}_t^2,
\label{jf}
\ee
where the operator ${\not\!\!D}_t$ is given in Eq.~(\ref{d}).
Eq.~(\ref{jf}) can be evaluated straightforwardly. Using
the identity, $\Tr\uv\Lambda\dmu A_\nu=(i/4)\Tr\uv Q\dmu Q\dnu Q$,
it leads to the topological term in the transformed action,
\eq
S_{\theta}={\sxyo\over8}\Tr\uv Q\dmu Q
\dnu Q,
\label{schiral}
\ee
where the bare coupling 
$\sxyo={1\over2\pi}(\pi+4\Delta\theta+\sin4\Delta\theta)$
is the {\it same} as the one obtained using nonabelian bosonization
and in the comment \cite{comment}.

TM used a seemingly different nonabelian chiral transformation,
$U_5^\prime(t)=\exp(itQ\phi\gamma^5)$, to directly bosonize Eq.~(\ref{sna}).
This leads to the transformed action $S^\prime=
\Tr{\bar \Psi}{\not\!\!D}_t^\prime\Psi+\ln{\cal J}_F^\prime$  
with
\eq 
{\not\!\!D}_t^\prime={\bf I}\otimes{\not\!\partial}+iU_5^\prime(t)
{\not\!\partial}U_5^\prime(t)-im\gamma^5e^{2iQ\phi(t-1)\gamma^5},
\label{dp}
\ee 
and
\eq 
\ln {\cal J}_F^\prime
=-{1\over2\pi}\int_0^1 dt(i\phi Q\gamma^5)({\not\!\!D}_t^\prime)^2 
\label{jfp}.
\ee 
We now prove that this is equivalent to the method used in Ref.~\cite{wang}
as described above. Notice that $U_5^\prime(t)=uU_5(t)u^\dagger$.
It is straightforward to show that
${\not\!\!D}_t^\prime=u{\not\!\!D}_t u^\dagger$. As a result,
the nonabelian chiral anomaly given by the Jacobians in
Eq.~(\ref{jf}) and Eq.~(\ref{jfp}) are identical, {\it i.e.}
$\ln{\cal J}_F=\ln{\cal J}_F^\prime$. Specifically,
$$
\ln {\cal J}_F=\ln{\cal J}_F^\prime= 
{\sxxo\over8}\Tr\dmu Q\dmu Q+{\sxyo\over8}\Tr\uv 
Q\dmu Q\dnu Q \label{sigma},
$$
where $\sxxo={1\over2\pi}\cos^22\Delta\theta$,
and $\sxyo={1\over2\pi}(\pi+4\Delta\theta+\sin4\Delta\theta)$ is
the same as the one given in Eq.~(\ref{schiral}).

\vspace{0.5truecm}
\bigskip
\noindent Ziqiang Wang and Vasiliki Plerou \\
Department of Physics \\
Boston College, Chestnut Hill, MA 02167


\vspace{-0.5truecm}
\begin{references}
\vspace{-0.8truecm}
\bibitem{comment} A. Tanaka and M. Machida, preceeding comment.
\bibitem{wang} Z. Wang, \pprl {\bf79}, 4002 (1997).
\bibitem{fujikawa} K. Fujikawa, Phys. Rev. D{\bf21}, 2848 (1980).
\bibitem{roskies} R. Roskies and F.~A. Schaposnik, Phys. Rev. D{\bf23},
558 (1981).
\bibitem{sch}R.~E. Gamboa Saravi, F.~A. Schaposnik, and J.~E. Solomin,
Nucl. Phys. B{\bf185}, 239 (1981).


\end{references}
\end{document}